\documentclass[preprint,aps,12pt,showpacs,nofootinbib,tightenlines]{revtex4}
\usepackage{mathrsfs}
\usepackage{amsmath}
\usepackage{amssymb}
\usepackage{epsfig}
\usepackage{graphicx}
\usepackage{multirow}
\newcommand{\psl}{ p \hspace{-2.4truemm}/ }

\newcommand{\esl}{ \epsilon \hspace{-1.5truemm}/ }

\def\be{\begin{eqnarray}}
\def\en{\end{eqnarray}}
\def\non{\nonumber\\}

\def\prd{{Phys. Rev. D}~}
\def\prl{{ Phys. Rev. Lett.}~}
\def\plb{{ Phys. Lett. B}~}

\def\epjc{{ Eur. Phys. J. C}~}

\begin{document}
\title{Probing $D^*_{s0}(2317)$ in the decays of $B$ to two charmed mesons}
\author{Zhi-Qing Zhang\footnote{e-mail: zhangzhiqing@haut.edu.cn.}, Meng-Ge Wang, Yan-Chao Zhao, Zhi-Lin Guan,  Na Wang}
\affiliation{\it \small $^1$  Department of Physics, Henan University of
Technology,\\ \small \it Zhengzhou, Henan 450052, P. R. China } 
\date{\today}
\begin{abstract}
We probe the inner structure of the meson $D^*_{s0}(2317)$ through the decays of
$B_{(s)}$ to two charmed mesons within pQCD approach. Assuming
$D^*_{s0}(2317)$ as a scalar meson with $\bar cs$ structure, we find
that the predictions for the branching ratios of the decays $B^+\to
D^{*+}_{s0}(2317)\bar D^{(*)0}, B^0\to D^{*+}_{s0}(2317)D^{(*)-}$
can explain data within errors. The branching ratios for the decays
$B_s\to D^{*+}_{s0}(2317)D_s^{(*)+}$ are estimated to reach up to
$10^{-3}$ order, which can be observed by the present LHCb and
SuperKEKB experiments. In this work, the decay constant of
the meson $D^*_{s0}(2317)$ is an input parameter. Unfortunately, its
value has been studied by many references but with large
uncertainties. Our calculation shows that a smaller decay constant of the meson $D^*_{s0}(2317)$ is
supported by compared with the present data, say $55\sim70$ MeV. We
also calculate the ratios $R_1=\frac{Br(B^+\to D^{*+}_{s0}(2317)\bar
D^0)}{Br(B^+\to D^{*+}_{s0}(2317)\bar D^{*0})})$ and
$R_2=\frac{Br(B^0\to D^{*+}_{s0}(2317)D^-)}{Br(B^0\to
D^{*+}_{s0}(2317)D^{*-})}$, which are valuable to determine the
inner structure of $D^*_{s0}(2317)$ by compared with the experimental results. Our predictions for the ratios $R_{1,2}$ are consistent
with the present data within errors.
 We expect that these two ratios can be well
measured by the future experiments through improving the measurement
accuracy for the decays $B^+\to D^{*+}_{s0}(2317)\bar D^{*0}$ and
$B^0\to D^{*+}_{s0}(2317)D^{*-}$.
\end{abstract}

 \vspace{1cm}

\maketitle

\section{Introduction}
The charmed-strange meson $D^*_{s0}(2317)$ was first observed by
BABAR Collaboration in the inclusive $D^+_s\pi^0$ invariant mass
distribution \cite{babar1}, and confirmed by CLEO \cite{cleo}. Then
BABAR and Belle collaborations probed the properties of this meson
through B meson to two charmed-meson decays \cite{belle,babar2,babar3}.
The branching ratios of these decays measured by BABAR and Belle were
averaged by the Particle Data Group (PDG) and given as \cite{pdg20}
\be
Br(B^+\to D^{*+}_{s0}(2317)(\to D^+_s\pi^0)\bar D^0)=(8.0^{+1.6}_{-1.3})\times10^{-4},\\
Br(B^+\to D^{*+}_{s0}(2317)(\to D^+_s\pi^0)\bar D^{*0})=(9\pm7)\times10^{-4},\\
Br(B^0\to D^{*+}_{s0}(2317)(\to D^+_s\pi^0)D^-)=(1.06\pm0.16)\times10^{-3},\\
Br(B^0\to D^{*+}_{s0}(2317)(\to
D^+_s\pi^0)D^{*-})=(1.5\pm0.6)\times10^{-3}. \en There have existed some
unsettled puzzles since this charmed-strange meson was observed in
2003: First, the low mass puzzle. Its measured mass is at least $150
MeV/c^2$ lower than the theoretical calculations from the potential
model \cite{god,kala}, lattice QCD \cite{bali0}. Second, the significantly large
branching ratio of the decay $D^{*-}_{s0}(2317)\to\pi^0D^-_s$ compared
with that of $D^{*-}_{s0}(2317)\to\gamma D^-_s$. BESIII measured that
$Br(D^{*-}_{s0}(2317)\to\pi^0D^-_s)=1.00^{+0.00}_{-0.14}\pm0.14$
\cite{bes3}, which differs from the expectation of the conventional
$\bar cs$ hypothesis. Third, uncertainties from the decay constant of the meson
$D^*_{s0}(2317)$. It has not been directly determined in experiment,
while the theoretical predictions coverd a very wide range (shown in
Table \ref{tab1}).
\begin{table}[!h]
\caption{The values of $f_{D^*_{s0}}$ (MeV) given
by different references. }
\begin{center}
\begin{tabular}{c|c|c|c|c|c|c|c}
\hline\hline  & Ref.\cite{colangelo} & Ref.\cite{jugeau}&Ref.\cite{herdoiza}& Ref.\cite{col}&Ref.\cite{hsieh}&Ref.\cite{bali}&Ref.\cite{segovia} \\
\hline
$f_{D^*_{s0}}$&$225\pm25$&$206\pm120$&$200\pm50$&$170\pm20$&$138\pm16$&$114^{+11.4}_{-10.2}$&$118.7$\\
\hline\hline
   &Ref.\cite{veseli}&Ref.\cite{verma}&Ref.\cite{cheng}&Ref.\cite{cheng}&Ref.\cite{cheng1}&Ref.\cite{faessler}&Ref.\cite{yaouanc} \\
\hline
$f_{D^*_{s0}}$&$110\pm18$&$74.4^{+10.4}_{-10.6}$&$71$&$60\pm13$&$67\pm13$&$67.1\pm4.5$&$44$\\
\hline\hline
\end{tabular}\label{tab1}
\end{center}
\end{table}
Because of these puzzles, the meson $D^{*}_{s0}(2317)$ attracts a lot of
attention. In order to solve these puzzles, many
various exotic explanations about its inner structure were proposed,
such as $DK$ molecule state \cite{barn,chenyq,guo,faes,guo1}, a
tetraquark state \cite{hycheng,tera,dmi,jrzhang}, or a mixture of a
$\bar c s$ state and a tetraquark state \cite{beve,mohler,liul}.
Certainly, its structure was also interpreted as a conventional $\bar
cs$ scalar meson in many references. For example, some people
considered that $D^*_{s0}(2317)$ is close to the threshold of $DK$,
so the low mass puzzle is because of the coupled-channel effects
\cite{van,hwang1,zhouzy,badalian}. Sometimes, the spontaneous
breaking of chiral symmetry was regarded as another possible
reason\cite{nowak,kolo}. Assuming the $D^*_{s0}(2317)$ as a
conventional charmed-strange meson, its properties were studied by
using constituent quark
model \cite{segovia}, covariant light-front approach \cite{cheng}, QCD sum rules \cite{colangelo,wei}, MIT bag model \cite{sadzi}, potential model
\cite{radford,fay,liujb}, Regge trajectories \cite{zhangal} and so on. More detailed discussion can be
found in Ref. \cite{chenhx}. Its pionic decay \cite{colangelo} and radiative decay \cite{wei}
 were researched in the light-cone QCD
sum rules, and obtained the results being consistent with data. The
productions of $D^*_{s0}(2317)$ in the $B_{(s)}$ decays
\cite{zhang,lirh,huangmq,zhaosm,segovia,shenyl,albert,chench,faus}
were also discussed.
In Ref. \cite{segovia}, the branching ratios of the decays $B\to
D^*_{s0}(2317)D^{(*)}$ were calculated in the factorization approximation
by using the constituent quark model. The authors found that
the meson $D^*_{s0}(2317)$ could be described as a conventional
$\bar c s$ state by introducing the finite $c$-quark mass effects.

In order to further reveal the inner structure of
$D^*_{s0}(2317)$, we intend to study the weak production of this
charmed-strange meson through the two charmed-meson $B_{(s)}$ decays,
some of which have been studied by using the light cone sum rules
(LCSR) \cite{lirh} and the relativistic quark model (RQM) \cite{faus}. In
layout of this paper is as follows. First, in Sec. II, we present
the analytic calculations about the $B_{(s)}$ decays to two
charmed mesons with $D^*_{s0}(2317)$ involved. Then, we give the
numerical results and discussions in Sec. III. A short summary of
our results is presented in the final part.
\section{The perturbative calculations}
In the pQCD approach, the only non-perturbative inputs are the light
cone distribution amplitudes (LCDAs) and the meson decay constants.
For the wave function of the heavy $B_{(s)}$ meson, we take
\cite{keum,ali} \be \Phi_{B_{(s)}}(x,b)= \frac{1}{\sqrt{2N_c}}
(\psl_{B_{(s)}} +m_{B_{(s)}}) \gamma_5 \phi_{B_{(s)}} (x,b).
\label{bmeson} \en Here only the contribution of Lorentz structure
$\phi_{B_{(s)}} (x,b)$ is taken into account, since the contribution
of the second Lorentz structure $\bar \phi_{B_{(s)}}$ is numerically
small \cite{cdlu} and has been neglected. For the distribution
amplitude $\phi_{B_{(s)}}(x,b)$ in Eq.(\ref{bmeson}), we adopt the
following model \be
\phi_{B_{(s)}}(x,b)=N_{B_{(s)}}x^2(1-x)^2\exp[-\frac{M^2_{B_{(s)}}x^2}{2\omega^2_b}-\frac{1}{2}(\omega_bb)^2],
\en where $\omega_b$ is a free parameter, we take
$\omega_b=0.4\pm0.04 (0.5\pm0.05)$ GeV for $B(B_s)$ in numerical
calculations, and $N_B=101.445(N_{B_s}=63.671)$ is the normalization
factor for $\omega_b=0.4(0.5)$. For $B_s$ meson, the SU(3) breaking
effects are taken into consideration.

The wave functions for the scalar meson $D^*_{s0}$ \footnote{From
now on, we will use $D^*_{s0}$ to denote $D^*_{s0}(2317)$ for simply
in some places.}, we use the form defined in Ref. \cite{chench} \be
\langle\bar D^{*+}_{s0}(2317)(p_2)|\bar
c_\beta(z)s_{\gamma}(0)|0\rangle&=&\frac{1}{\sqrt{2N_c}}\int
dxe^{ip_2\cdot
z}\left[(\psl_{2})_{lj}+m_{D^*_{s0}}I_{lj}\right]\phi_{D^*_{s0}}.
\en It is noticed that the distribution amplitudes which associate
with the nonlocal operators $\bar c(z)\gamma_\mu s$ and $\bar c(z)s$
are different. The difference between them is order of $\bar
\Lambda/m_{D^*_{s0}}\sim(m_{D^*_{s0}}-m_c)/m_{D^*_{s0}}$. If we set
$m_{D^*_{s0}}\sim m_c$, we can get these two distribution amplitudes
being very similar. For the leading power calculation, it is
reasonable to parameterize them in the same form as \be
\phi_{D^*_{s0}}(x)=\tilde{f}_{D^*_{s0}}6x(1-x)\left[1+a(1-2x)\right]
\en in the heavy quark limit. Here the decay constant
$\tilde{f}_{D^*_{s0}}$ is defined through the matrix element of the
scalar current \be \langle0|\bar
sc|D^*_{s0}(\textbf{p})\rangle=\tilde{f}_{D^*_{s0}}m_{D^*_{s0}} \label{barfDst0} \en and the shape parameter $a=-0.21$
\cite{lirh} is fixed under the condition that the distribution
amplitude $\phi_{D^*_{s0}}(x)$ possesses the maximum at $
x=m_c/m_{D^*_{s0}}$ with $m_c=1.275$ GeV. It is
worthwhile to point out that the intrinsic $b$ dependence of this
charmed meson's wave function has been neglected in our analysis.

As for the wave functions of the mesons $D^{(*)}$, we use the form derived
in Ref. \cite{kurimoto} \be
\int\frac{d^4\omega}{(2\pi)^4}e^{ik\cdot\omega}\langle0|\bar c_\beta(0)u_{\gamma}(\omega)|\bar D^0\rangle&=&-\frac{i}{\sqrt{2N_c}}[(\psl_D+m_D)\gamma_5]_{\gamma\beta}\phi_D(x,b),\\
\int\frac{d^4\omega}{(2\pi)^4}e^{ik\cdot\omega}\langle0|\bar
c_\beta(0)u_{\gamma}(\omega)|\bar
D^{*0}\rangle&=&-\frac{i}{\sqrt{2N_c}}[(\psl_{D^*}+m_{D^*})\esl_L]_{\gamma\beta}\phi^L_{D^*}(x,b),
\en where $\esl_L$ is the longitudinal polarization vector. In this
work only the longitudinal polarization component is used. Here we
take the best-fitted form $\phi_D^{(*)}$ from B to charmed meson
decays derived in \cite{lirh1} as \be
\phi_{D}(x,b)=\frac{f_{D}}{2\sqrt{2N_c}}6x(1-x)[1+C_{D}(1-2x)]\exp[\frac{-\omega^2b^2}{2}].
\en For the wave function $\phi_{D_{s}}(x,b)$, it has the similar
expression as $\phi_D(x,b)$ except with the different parameters. These prameters are
given as follows: $f_D=223$ MeV, $f_{D_s}=274$ MeV, and
$C_{D_{(s)}}=0.5$ $(0.4)$, $\omega_{D_{(s)}}=0.1(0.2)$ GeV
\cite{lirh1}. For the wave function $\phi_{D^*_{(s)}}(x,b)$, we take
the same distribution amplitude with that of the pseudoscalar meson
$D_{(s)}$ because of their small mass difference. The decay constants $f_{D^*}$ and $f_{D^*_s}$ are given by the
relations \be
f_{D^{*-}}=\sqrt{\frac{M_{D^-}}{M_{D^{*-}}}}f_{D^-},\;\;\;
f_{D^{*-}_s}=\sqrt{\frac{M_{D^-_s}}{M_{D^{*-}_s}}}f_{D^-_s}. \en.

For these processes considered, the weak effective Hamiltonian
$H_{eff}$ can be written as: \be
H_{eff}&=&\frac{G_F}{\sqrt{2}}\left\{\sum_{q=u,c}V_{qb}V^*_{qD}[C_1(\mu)O^q_1(\mu)+C_2(\mu)O^q_2(\mu)]\right.\non
&&\left.-V_{tb}V^*_{tD}
\left[\sum^{10}_{i=3}C_i(\mu)O_i(\mu)\right]\right\}+H.C., \en where
$V_{qb(D)}$ and $V_{tb(D)}$ with $D=d,s$ are CKM matrix elements.
The local four-quark operators $O_i(i=1,...,10)$ include three type
operators: current-current operators ($O^q_{1,2}$), QCD
penguin ($O_{3\sim6}$) and electroweak penguin
opertors ($O_{7\sim10}$), \be O^q_1&=&(\bar
q_{\alpha}b_{\beta})_{V-A}(\bar D_{\beta}q_{\alpha})_{V-A},\;\;\;
O^q_2=(\bar q_{\alpha}b_{\alpha})_{V-A}(\bar
D_{\beta}q_{\beta})_{V-A},\\ \non
O_3&=&(\bar D_\alpha b_\alpha)_{V-A}\sum_{q'}(\bar q'_\beta q'_\beta)_{V-A},\;\;\; O_4=(\bar D_{\beta}b_{\alpha})_{V-A}\sum_{q'}(\bar q'_\alpha q'_\beta)_{V-A},\\
O_5&=&(\bar D_\alpha b_\alpha)_{V-A}\sum_{q'}(\bar q'_\beta q'_\beta)_{V+A},\;\;\; O_6=(\bar D_{\beta}b_{\alpha})_{V-A}\sum_{q'}(\bar q'_\alpha q'_\beta)_{V+A},\\
O_7&=&\frac{3}{2}(\bar D_\alpha b_\alpha)_{V-A}\sum_{q'}e_{q'}(\bar q'_\beta q'_\beta)_{V+A},\;\;\; O_8=\frac{3}{2}(\bar D_{\beta}b_{\alpha})_{V-A}\sum_{q'}e_{q'}(\bar q'_\alpha q'_\beta)_{V+A},\\
O_9&=&\frac{3}{2}(\bar D_\alpha b_\alpha)_{V-A}\sum_{q'}e_{q'}(\bar
q'_\beta q'_\beta)_{V-A},\;\;\; O_{10}=\frac{3}{2}(\bar
D_{\beta}b_{\alpha})_{V-A}\sum_{q'}e_{q'}(\bar q'_\alpha
q'_\beta)_{V-A}, \en where $(\bar q'_\alpha q'_\beta)_{V\pm A}=\bar
q'_\alpha \gamma_{\nu}(1\pm\gamma_5) q'_\beta$ with $\alpha,\beta$
being the color indices and $q'$ represent the active quarks at the $m_b$
scale, which can be $u,d,s,c$ and $b$.
\begin{figure}[t]
\vspace{-4cm} \centerline{\epsfxsize=18 cm \epsffile{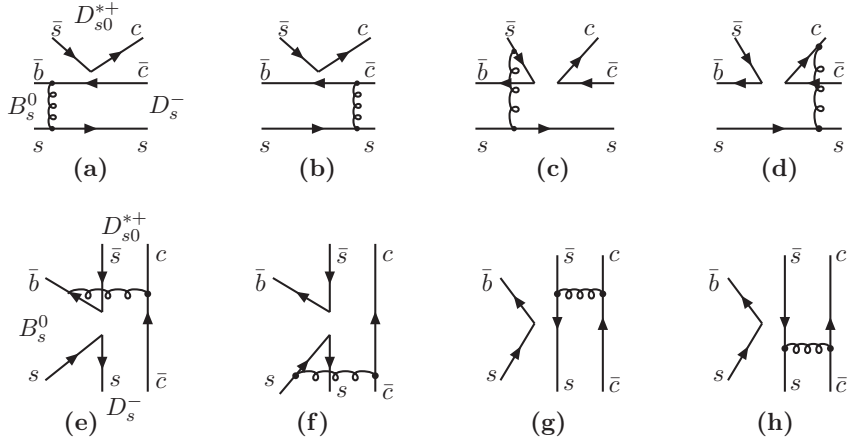}}
\vspace{-15.5cm} \caption{ A part of diagrams contributing to the
$ B^0_s\to D^{*+}_{s0}(2317)D^-_s$ decay.}
 \label{Figure1}
\end{figure}
We calculate in the light-cone coordinate, where a vector is defined
as \be
P_{\mu}=(\frac{P_0+P_3}{\sqrt{2}},\frac{P_0-P_3}{\sqrt{2}},P_1,P_2).
\en When working in the rest frame of $B_{(s)}$ meson and defining
the direction where $D^{*}_{s0}$ moves as the positive direction of
$z$-axis, we can write the momenta of $B, D^{*}_{s0}, D_s$
mesons as \be P_{B}=\frac{m_{B}}{\sqrt2}(1,1,\textbf{0}_\perp),
P_{D^{*}_{s0}}=\frac{m_{B}}{\sqrt2}(1-r^2_{D_s},r^2_{D^{*}_{s0}},\textbf{0}_\perp),
P_{D_s}=\frac{m_{B}}{\sqrt2}(r^2_{D_s},1-r^2_{D^{*}_{s0}},\textbf{0}_\perp),
\en where $r_{D^*_{s0}}=m_{D^*_{s0}}/m_{B},r_{D_s}=m_{D_s}/m_{B}$
and $\textbf{0}_\perp$ is zero two-component vector. If using
$k_1,k_2,$ and $k_3$ to denote the momenta carried by the light
quark in $B$ and two charmed mesons, we have \be
k_1=(\frac{m_B}{\sqrt2}x_1,0,\textbf{k}_{1\perp}),k_2=(\frac{m_B}{\sqrt2}(1-r^2_{D_s})x_2,0,\textbf{k}_{2\perp}),
k_3=(0,\frac{m_B}{\sqrt2}(1-r^2_{D^{*}_{s0}})x_3,\textbf{k}_{3\perp}).
\en

Here we consider the decays $B^0_s\to D^{*+}_{s0}(2317)D^-_s$ and
$B^0_s\to D^{*-}_{s0}(2317)D^+_s$ as examples, which will include all types of
Feynman diagram amplitudes we need. For the decay $B^0_s\to
D^{*+}_{s0}(2317)D^-_s$, we give a part of its Feynman diagrams at
leading order in Fig.1, where the scalar meson is in the emission
(upper) position. If changing the positions of $D^{*+}_{s0}(2317)$
and $D^-_s$ for the annihilation type Feynman diagrams (the second
line in Fig.1), we will obtain another part of Feynman diagrams
which can also contribute to the decay $B^0_s\to
D^{*+}_{s0}(2317)D^-_s$. These Feynman diagrams are given in Fig.
2. The Feynman diagrams for the decay $B^0_s\to
D^{*-}_{s0}(2317)D^+_s$ are totally same with those for the decay
$B^0_s\to D^{*+}_{s0}(2317)D^-_s$ and can be obtained just by
changing $D^{*+}_{s0}(D^-_s)$ to $D^+_s(D^{*-}_{s0})$ in Fig.1 and 2. There are two
points we need to emphasize: (I) Besides of the different positions
for the final states in the annihilation type Feynman diagrams
between Fig.1 and 2, another difference is that the former
with $c\bar c$ pair generated from a hard gluon, while the later
with $s\bar s$ pair generated. (II) In order to distinguish these
amplitudes for the decays $B^0_s\to D^{*+}_{s0}(2317)D^-_s$ and
$B^0_s\to D^{*-}_{s0}(2317)D^+_s$ from each other, we add the
character "c" in the subscript for each amplitude which corresponds
to the Feynman diagram with a conventional charmed meson being the
emission (upper) position. For the amplitudes corresponding to the
annihilation diagrams with $s\bar s$ pair generated from a hard
gluon, we add another character "s" in the subscripts. In the
following, we give the detail expressions of the Feynman diagram amplitudes for the decay $B^0_s\to
D^{*+}_{s0}(2317)D^-_s$. Fig.1(a) and 1(b) are the factorization
emission diagrams, Fig.1(c) and 1(d) are the nonfactorization
emission ones, the corresponding amplitudes can be written as \be
\mathcal{F}^{LL}_{e}&=&8\pi C_FM^4_{B_s}f_{D^*_{s0}}\int_0^1 d x_{1}
dx_{3}\, \int_{0}^{\infty} b_1 db_1 b_3 db_3\,
\phi_{B_s}(x_1,b_1)\phi_{D_s}(x_3)\non && \times
[1+r_{D_s}+(1-2r_{D_s})x_3]
E_e(t^{(1)}_e)S_t(x_3)h_e(x_1,x_3(1-r^2_{D^*_{s0}}),b_1,b_3)\non &&
+[2(r_c+1)r_{D_s}-r_c-r_{D_s}^2]E_e(t^{(2)}_e)S_t(x_1)
h_e(x_3,x_1(1-r^2_{D^*_{s0}}),b_3,b_1)], \label{fe1} \\
\mathcal{F}^{SP}_{e}&=&16\pi
C_FM^4_{B_s}f_{D^*_{s0}}r_{D^*_{s0}}\int_0^1 d x_{1} dx_{3}\,
\int_{0}^{\infty} b_1 db_1 b_3 db_3\,
\phi_{B_s}(x_1,b_1)\phi_{D_s}(x_3)\non && \times
[1+r_{D_s}(2+r_{D_s}+x_3(1-4r_{D_s}))]
E_e(t^{(1)}_e)S_t(x_3)h_e(x_1,x_3(1-r^2_{D^*_{s0}}),b_1,b_3)\non &&
+[-r_c(1-4r_{D_s})+2r_{D_s}(1-r_{D_s})]E_e(t^{(2)}_e)S_t(x_1)
h_e(x_3,x_1(1-r^2_{D^*_{s0}}),b_3,b_1)], \label{fe1}\\
\mathcal{M}^{LL}_{e}&=&32\pi C_f m_{B_s}^4/\sqrt{2N_C}\int_0^1 d
x_{1} dx_{2} dx_{3}\, \int_{0}^{\infty} b_1 db_1 b_2
db_2\,\phi_{B_s}(x_1,b_1)\phi_{D_s}(x_2)\phi_{D^*_{s0}}(x_3)\non &&
\times\left\{\left[x_2-r_{D_s}x_3(1-2r_{D_s})\right]E_{en}(t^{(1)}_{en})h^{(1)}_{en}(x_1,x_2,x_3,b_1,b_2)
\right.\non &&\left.
 +\left[x_2-1-(1-r_{D_s})x_3+r_cr_{D^*_{s0}}\right]E_{en}(t^{{2}}_{en})h^{(2)}_{en}(x_1,x_2,x_3,b_1,b_2)\right\}, \label{nfe1}\\
\mathcal{M}^{LR}_{e}&=&32\pi C_f m_{B_s}^4/\sqrt{2N_C}\int_0^1 d
x_{1} dx_{2} dx_{3}\, \int_{0}^{\infty} b_1 db_1 b_2
db_2\,\phi_{B_s}(x_1,b_1)\phi_{D_s}(x_2)\phi_{D^*_{s0}}(x_3)\non &&
\times(1+r_{D_s})\left\{r_{D^*_{s0}}\left[x_2+r_{D_s}(1+r_{D_s})x_3\right]E_{en}(t^{(1)}_{en})h^{(1)}_{en}(x_1,x_2,x_3,b_1,b_2)
\right.\non &&\left.
 -\left[r_c+r_{D^*_{s0}}(1-x_2)+r_{D_s}(r_{D_s}+1)r_{D^*_{s0}}x_3\right]E_{en}(t^{(2)}_{en})h^{(2)}_{en}(x_1,x_2,x_3,b_1,b_2)\right\},\;\;\;\; \label{nfe1}
 \en \be
\mathcal{M}^{SP}_{e}&=&32\pi C_f m_{B_s}^4/\sqrt{2N_C}\int_0^1 d
x_{1} dx_{2} dx_{3}\, \int_{0}^{\infty} b_1 db_1 b_2
db_2\,\phi_{B_s}(x_1,b_1)\phi_{D_s}(x_2)\phi_{D^*_{s0}}(x_3)\non &&
\times\left\{\left[x_2-(r_{D^*_{s0}}-1)x_3\right]E_{en}(t^{(1)}_{en})h^{(1)}_{en}(x_1,x_2,x_3,b_1,b_2)
\right.\non &&\left.
-\left[1-x_2-r_{D_s}x_3-r_cr_{D^*_{s0}}\right]E_{en}(t^{(2)}_{en})h^{(2)}_{en}(x_1,x_2,x_3,b_1,b_2)\right\},
\label{nfe1} \en where $r_{D^*_{s0}}=m_{D^*_{s0}}/m_{B_s},
r_{D_s}=m_{D_s}/m_{B_s},r_{c}=m_c/m_{B_s}$ and $f_{D^*_{s0}}$ is the decay
constant of the scalar meson $D^*_{s0}(2317)$. As we know that the double logarithms $\alpha_sln^2x$
produced by the radiative corrections are not small expansion
parameters when the end point region is important, in order to
improve the perturbative expansion, the threshold resummation of
these logarithms to all order is needed, which leads to a quark jet
function \be
S_t(x)=\frac{2^{1+2c}\Gamma(3/2+c)}{\sqrt{\pi}\Gamma(1+c)}[x(1-x)]^c,
\en with $c=0.35$. It is effective to smear the end point
singularity with a momentum fraction $x\to0$. This factor will also
appear in the factorizable annihilation amplitudes.

As to the (non)factorizable annihilation amplitudes for the second
line Feynman diagrams in Fig.1 can be obtained by the Feynman rules
and are given as \be \mathcal{M}^{LL}_{an}&=&32\pi C_f
m_{B_s}^4/\sqrt{2N_C}\int_0^1 d x_{1} dx_{2}
dx_{3}\,\int_{0}^{\infty} b_1 db_1 b_3 db_3\,
\phi_{B_s}(x_1,b_1)\phi_{D^*_{s0}}(x_2)\phi_{D_s}(x_3)\non &&
\times\left\{\left[x_2-1+r_{D_s}r_{D^*_{s0}}(x_2+x_3-4)\right]
E_{an}(t^{(1)}_{an})h^{(1)}_{an}(x_1,x_2,x_3,b_1,b_3)\right.\non &&\left.+\left[1-x_3-r_{D_s}r_{D^*_{s0}}(x_2+x_3-2)\right]E_{an}(t^{(2)}_{an})h^{(2)}_{an}(x_1,x_2,x_3,b_1,b_3)\right\},\\
\mathcal{M}^{LR}_{an}&=&32\pi C_f m_{B_s}^4/\sqrt{2N_C}\int_0^1 d
x_{1} dx_{2} dx_{3}\,\int_{0}^{\infty} b_1 db_1 b_3 db_3\,
\phi_{B_s}(x_1,b_1)\phi_{D^*_{s0}}(x_2)\phi_{D_s}(x_3)\non &&
\times\left\{\left[r_{D^*_{s0}}(x_2+1)-r_{D_s}(x_3+1)\right]
E_{an}(t^{(1)}_{an})h^{(1)}_{an}(x_1,x_2,x_3,b_1,b_3)\right.\non
&&\left.-\left[r_{D_s}(1-x_3)+r_{D^*_{s0}}(x_2-1)\right]E_{an}(t^{(2)}_{an})h^{(2)}_{an}(x_1,x_2,x_3,b_1,b_3)\right\},
\en \be \mathcal{M}^{SP}_{an}&=&32\pi C_f
m_{B_s}^4/\sqrt{2N_C}\int_0^1 d x_{1} dx_{2}
dx_{3}\,\int_{0}^{\infty} b_1 db_1 b_3 db_3\,
\phi_{B_s}(x_1,b_1)\phi_{D^*_{s0}}(x_2)\phi_{D_s}(x_3)\non &&
\times\left\{\left[1-x_3-r_{D_s}r_{D^*_{s0}}(x_2+x_3-4)\right]
E_{an}(t^{(1)}_{an})h^{(1)}_{an}(x_1,x_2,x_3,b_1,b_3)\right.\non
&&\left.-\left[1-x_2-r_{D_s}r_{D^*_{s0}}(x_2+x_3-2)\right]E_{an}(t^{(2)}_{an})h^{(2)}_{an}(x_1,x_2,x_3,b_1,b_3)\right\},
\en \be \mathcal{F}^{LL}_{a}&=&-\mathcal{F}^{LR}_{a}=8\pi C_f
f_{B_s}m_{B_s}^4\int_0^1 d x_{2} dx_{3}\, \int_{0}^{\infty} b_2 db_2
b_3 db_3\, \phi_{D^*_{s0}}(x_2)\phi_{D_s}(x_3)\non
&&\left\{\left[x_3-1+2r_{D_s}r_{D^*_{s0}}(x_3-2)\right]
E_{a}(t^{(1)}_a)S_t(x_3)h_{a}(\overline{(1-r^2_{D_s})x_2},\overline{(1-r_{D^*_{s0}}^2)x_3},b_2,b_3)
\right.\non &&\left.+ \left[1-x_2-2r_{D_s}r_{D^*_{s0}}(x_2-2)\right]E_{a}(t^{(2)}_a)S_t(x_2)h_{a}(\overline{(1-r_{D^*_{s0}}^2)x_3},\overline{(1-r^2_{D_s})x_2},b_3,b_2)\right\},\non\\
\mathcal{F}^{SP}_{a}&=&16\pi C_f f_{B_s}m_{B_s}^4\int_0^1 d x_{2}
dx_{3}\, \int_{0}^{\infty} b_2 db_2 b_3 db_3\,
\phi_{D^*_{s0}}(x_2)\phi_{D_s}(x_3)\non
&&\left\{\left[2r_{D^*_{s0}}+r_{D_s}(1-x_3)\right]
E_{a}(t^{(1)}_a)S_t(x_3)h_{a}(\overline{(1-r^2_{D_s})x_2},\overline{(1-r_{D^*_{s0}}^2)x_3},b_2,b_3)
\right.\non &&\left.+
\left[2r_{D_s}+r_{D^*_{s0}}(1-x_2)\right]E_{a}(t^{(2)}_a)S_t(x_2)h_{a}(\overline{(1-r_{D^*_{s0}}^2)x_3},\overline{(1-r^2_{D_s})x_2},b_3,b_2)\right\},
\en
where $\overline{(1-r^2_{D_s})x_2}=1-(1-r^2_{D_s})x_2, \overline{(1-r_{D^*_{s0}}^2)x_3}=1-(1-r_{D^*_{s0}}^2)x_3$. The hard scales,
evolution factors, the expressions of the Sudakov factors and the
functions of the hard kernels in the above amplitudes can be found in
appendix A.
\begin{figure}[t]
\vspace{-4cm} \centerline{\epsfxsize=18 cm
\epsffile{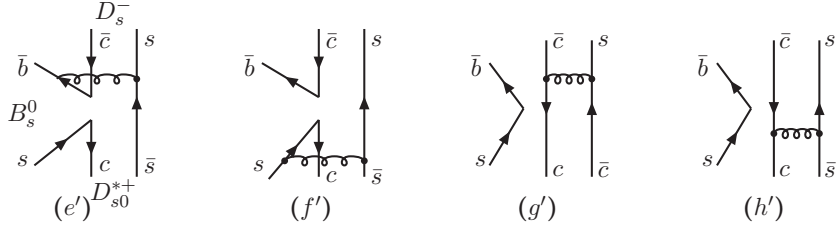}} \vspace{-18.5cm} \caption{ Another
part of diagrams contributing to the $B^0_s\to
D^{*+}_{s0}(2317)D^-_s$ decay.}
 \label{Figure2}
\end{figure}

Another type of annihilation Feynman diagrams contributing to the decay $ B_s\to
D^{*+}_{s0}(2317)D^-_s$ are shown in Fig.2, and the
corresponding amplitudes are written as \be
\mathcal{M}^{LL}_{ancs}&=&-32\pi C_f m_{B_s}^4/\sqrt{2N_C}\int_0^1 d
x_{1} dx_{2} dx_{3}\,\int_{0}^{\infty} b_1 db_1 b_2 db_2\,
\phi_{B_s}(x_1,b_1)\phi_{D^*_{s0}}(x_2)\phi_{{D_s}}(x_3)\non &&
\times\left\{\left[x_3-r_{D_s}r_{D^*_{s0}}(x_2+x_3+2)\right]
E_{an}(t^{(1)}_{ancs})h^{(1)}_{ancs}(x_1,x_3,x_2,b_1,b_2)\right.\non &&\left.-\left[x_2-r_{D_s}r_{D^*_{S0}}(x_2+x_3)\right]E_{an}(t^{(2)}_{ancs})h^{(2)}_{ancs}(x_1,x_3,x_2,b_1,b_2)\right\},\\
\mathcal{M}^{LR}_{ancs}&=&-32\pi C_f m_{B_s}^4/\sqrt{2N_C}\int_0^1 d
x_{1} dx_{2} dx_{3}\,\int_{0}^{\infty} b_1 db_1 b_2 db_2\,
\phi_{B_s}(x_1,b_1)\phi_{D^*_{s0}}(x_2)\phi_{D_s}(x_3)\non &&
\times\left\{\left[r_{D_s}(2-x_3)-r_{D^*_{s0}}(x_2-2)\right]
E_{an}(t^{(1)}_{ancs})h^{(1)}_{ancs}(x_1,x_3,x_2,b_1,b_2)\right.\non &&\left.+\left[r_{D^*_{s0}}x_2+r_{D_s}x_3\right]E_{an}(t^{(2)}_{ancs})h^{(2)}_{ancs}(x_1,x_3,x_2,b_1,b_2)\right\},
\en
\be
\mathcal{M}^{SP}_{ancs}&=&32\pi C_f m_{B_s}^4/\sqrt{2N_C}\int_0^1 d
x_{1} dx_{2} dx_{3}\,\int_{0}^{\infty} b_1 db_1 b_2 db_2\,
\phi_{B_s}(x_1,b_1)\phi_{D^*_{s0}}(x_2)\phi_{D_s}(x_3)\non &&
\times\left\{\left[x_2-r_{D_s}r_{D^*_{s0}}(x_2+x_3+2)\right]
E_{an}(t^{(1)}_{ancs})h^{(1)}_{ancs}(x_1,x_3,x_2,b_1,b_2)\right.\non
&&\left.-\left[x_3-r_{D_s}r_{D^*_{s0}}(x_2+x_3)\right]E_{an}(t^{(2)}_{ancs})h^{(2)}_{ancs}(x_1,x_3,x_2,b_1,b_2)\right\},
\en \be \mathcal{F}^{LL}_{acs}&=&-\mathcal{F}^{LR}_{acs}=-8\pi C_f
f_{B_s}m_{B_s}^4\int_0^1 d x_{2} dx_{3}\, \int_{0}^{\infty} b_2 db_2
b_3 db_3\, \phi_{D^*_{s0}}(x_2)\phi_{D_s}(x_3)\non
&&\left\{\left[1-x_2+2r_{D_s}r_{D^*_{s0}}(x_2-2))\right]
E_{a}(t^{(1)}_{acs})S_t(x_2)h_{a}(x_3,(1-r_{D^*_{s0}}^2-r^2_{D_s})x_2,b_3,b_2)
\right.\non &&\left.-\left[1-x_3+2r_{D_s}r_{D^*_{s0}}(x_3-2)\right]E_{a}(t^{(2)}_{acs})S_t(x_3)h_{a}(x_2,(1-r_{D^*_{S0}}^2-r^2_{D_s})x_3,b_2,b_3))\right\},\label{ckms4}\non\\
\mathcal{F}^{SP}_{acs}&=&16\pi C_f f_{B_s}m_{B_s}^4\int_0^1 d x_{2}
dx_{3}\, \int_{0}^{\infty} b_2 db_2 b_3 db_3\,
\phi_{D^*_{s0}}(x_2)\phi_{D_s}(x_3)\non
&&\left\{\left[-2r_{D_s}+r_{D^*_{s0}}(1-x_2)+r_c\right]
E_{a}(t^{(1)}_{acs})S_t(x_2)h_{a}(x_3,(1-r_{D^*_{s0}}^2-r^2_{D_s})x_2,b_3,b_2)
\right.\non &&\left.+
\left[2r_{D^*_{s0}}-r_{D_s}(1-x_3)-r_c\right]E_{a}(t^{(2)}_{acs})S_t(x_3)h_{a}(x_2,(1-r_{D^*_{s0}}^2-r^2_{D_s})x_3,b_2,b_3))\right\}.\non
\en The amplitudes for the decay $B_s\to D^{*-}_{s0}D^+_s$ are
listed in Appendix B. As to the decays with a vector meson
$D^*_{(s)}$ involved, the corresponding amplitudes can be also
obtained from the Feymann rules, which are not listed for
simplicity. Combining these amplitudes, we can obtain the total
decay amplitude of each considered channel \be \mathcal{A}(B^+\to
D^{*+}_{s0}D^{0})&=&\frac{G_F}{\sqrt2}\left\{V^*_{cb}V_{cs}\left[\emph{F}^{LL}_{e}(a_1)+\emph{M}^{LL}_{en}(C_1)\right]
+V_{ub}V_{us}\left[\emph{F}^{LL}_{a}(a_1)+\emph{M}^{LL}_{an}(C_1)\right]\right.\non
&& \left.-V^*_{tb}V_{ts}\left[\emph{F}^{LL}_{e}(a_4+a_{10})
+\emph{M}^{LL}_{en}(C_3+c_9)+\emph{F}^{SP}_{e}(a_6+a_8)\right.\right.\non
&&
\left.\left.+\emph{M}^{LR}_{en}(C_5+C_7)+\emph{F}^{LL}_{a}(a_4+a_{10})+\emph{F}^{SP}_{a}(a_6+a_8)
\right.\right.\non &&
\left.\left.+\emph{M}^{LL}_{an}(C_3+C_9)+\emph{M}^{LR}_{an}(C_5+C_7)\right]\right\},
\label{dspi}\\
\mathcal{A}(B^0\to
D^{*+}_{s0}D^{-})&=&\frac{G_F}{\sqrt2}\left\{V^*_{cb}V_{cs}[\emph{F}^{LL}_{e}(a_1)+\emph{M}^{LL}_{en}(C_1)]-V^*_{tb}V_{ts}\left[
\emph{F}^{LL}_{e}(a_4+a_{10})\right.\right.\non
&&\left.\left.+\emph{F}^{SP}_{e}(a_6+a_8)+\emph{M}^{LL}_{en}(C_3+C_9)+\emph{M}^{LR}_{en}(C_5+C_7)
+\emph{F}^{LL}_{a}(a_4-\frac{a_{10}}{2})\right.\right.\non && \left.\left.+\emph{F}^{SP}_{a}(a_6-\frac{a_8}{2})+\emph{M}^{LL}_{an}(C_3-\frac{C_9}{2})+\emph{M}^{LR}_{an}(C_5-\frac{C_7}{2})\right]\right\},
\en\be
\emph{A}(B^0_s\to
D^{*+}_{s0}D^-_s)&=&\frac{G_F}{\sqrt2}\left\{V^*_{cb}V_{cs}\left[\emph{F}^{LL}_{e}(a_1)+M^{LL}_{en}(C_1)+F^{LL}_{acs}(a_2)
+M^{LL}_{ancs}(C_2)\right]-V^*_{tb}V_{ts}\left[\right.\right.\non &&
\left.\left.F^{LL}_{e}(a_4+a_{10})+F^{SP}_{e}(a_6+a_{8})+M^{LL}_{en}(C_3+C_9)+M^{LR}_{en}(C_5+C_7)\right.
\right.\non &&
\left.\left.+F^{LL}_{acs}(a_3-a_5-a_7+a_9)+M^{LL}_{ancs}(C_4+C_{10})+M^{SP}_{ancs}(C_6+C_8)\right.\right.\non
&&\left.\left.
F^{LL}_{a}(a_3+a_4-a_5+\frac{a_7}{2}-\frac{a_9}{2}-\frac{a_{10}}{2})+M^{LL}_{an}(C_3+C_4-\frac{C_9}{2}-\frac{10}{2})
\right.\right.\non && \left.\left.+F^{SP}_{a}(a_6-\frac{a_8}{2})+M^{SP}_{an}(C_6-\frac{C_8}{2})+M^{LR}_{an}(C_5-\frac{C_7}{2})\right]\right\},\\
\emph{A}(B^0_s\to
D^{*-}_{s0}D^+)&=&\frac{G_F}{\sqrt2}\left\{V^*_{cb}V_{cd}\left[F^{LL}_{ec}(a_1)+M^{LL}_{enc}(C_1)\right]
-V^*_{tb}V_{td}\left[F^{LL}_{ec}(a_4+a_{10})\right.\right.\non
&&\left.\left.+M^{LL}_{enc}(C_3+C_9)+M^{LR}_{enc}(C_5+C_7)+F^{SP}_{ec}(a_6+a_8)
+F^{LL}_{ac}(a_4-\frac{a_{10}}{2})\right.\right.\non
&&\left.\left.+M^{LL}_{anc}(C_3-\frac{C_9}{2})+F^{SP}_{ac}(a_6-\frac{a_8}{2})
+M^{LR}_{anc}(C_5-\frac{C_7}{2})\right]\right\}. \en
The
amplitudes for the decay $B^0_s\to D^{*-}_{s0}D^+_s$ can be obtained
from those for the decay $B^0_s \to D^{*+}_{s0}D^-_s$ by deleting
(adding) the character "c" from (to) the subscript of each amplitude where there is (not) a character "c".
\section{the Numerical results and discussions}
We use the following input parameters in the numerical calculations
\cite{pdg20,keum}: \be
f_B&=&210 MeV, f_{B_s}=230 MeV, M_B=5.28 GeV, M_{B_s}=5.37 GeV, \\
\tau_B^\pm&=&1.638\times 10^{-12} s,\tau_{B^0}=1.519\times 10^{-12} s, \tau_{B_s}=1.512\times 10^{-12} s,\\
M_{W}&=&80.38 GeV, M_{D^*_{s0}}=2.3178 GeV.\en For the CKM matrix
elements,
we adopt the Wolfenstein parametrization and the updated values
$A=0.790^{+0.017}_{-0.012}, \lambda=0.22650\pm0.00048, \bar\rho=0.141^{+0.016}_{-0.017}$
and $\bar\eta=0.357\pm0.011$ \cite{pdg20}.

Generally speaking that the branching ratio of the charged channel
should not be less than that of the corresponding neutral one. For example,
Particle Data Group(PDG) gives that $Br(B^+\to D^+_s\bar D^0)=(9.0\pm0.9)\times10^{-3}$,
which is larger than $Br(B^0\to
D^+_sD^-)=(7.2\pm0.8)\times10^{-3}$ \cite{pdg20}. Similarly our
calculations (given in Table II) also show that the branching ratio
of the charged decay $B^+\to D^+_s(2317)\bar D^{(*)0}$ is slightly
larger than that of the neutral decay $B^0\to D^+_s(2317)D^{(*)-}$.
But data are just opposite \cite{pdg20}. Certainly, there still exist
large errors in the experimental results, especially for the branching ratios of the decays with a vector
meson $D^*$ involved. We expect more accurate experimental
results in the future LHCb and Super KEKB experiments.
Theoretically, the decays $B^+\to D^+_s(2317)\bar
D^{(*)0}$ and $B^0\to D^+_s(2317)D^{(*)-}$ have the same CKM matrix
elements and Wilson coefficients for the factorizable and
nonfactorizable emission amplitudes, it is different in the
amplitudes from the annihilation diagrams, while their contributions are small
which will be discussed later. Furthermore, there exist similar
transition form factors for isospin symmetry between each pair of
decay channels. So they should have similar branching ratios.

\begin{table}
\caption{Branching ratios ($\times10^{-4}$) of the decays $B^+\to D^{*+}_{s0}(2317)\bar D^{(*)0}$ and
$B^0\to D^{*+}_{s0}(2317)D^{(*)-}$ with
different values $f_{D^*_{s0}}=55, 60, 67$ MeV, where the
errors for these entries correspond to the uncertainties in the
$w_{b}=0.4\pm0.04$ GeV for $B$ meson, the hard scale
$t$ varying from $0.75t$ to $1.25t$, and the CKM matrix elements.}
\begin{center}
\begin{tabular}{c|c|c|c|c}
\hline\hline Modes & $f_{D^*_{s0}}=55$ &  $f_{D^*_{s0}}=60$ & $f_{D^*_{s0}}=67$ & Data \cite{pdg20}  \\
\hline
$Br(B^+\to D^{*+}_{s0}(2317)\bar D^0)$&$7.5^{+3.3+0.1+0.3}_{-2.2-0.3-0.3}$&$8.9^{+4.0+0.5+0.4}_{-2.6-0.2-0.3}$&$11.2^{+4.0+0.3+0.4}_{-2.8-0.2-0.4}$&${8.0^{+1.6}_{-1.3}}$\\
$Br(B^+\to D^{*+}_{s0}(2317)\bar D^{*0})$ &$12.0^{+4.9+1.9+0.5}_{-3.5-1.2-0.4}$ &$14.4^{+6.0+2.3+0.5}_{-4.3-1.5-0.4}$&$18.3^{+7.1+2.7+0.7}_{-5.4-1.7-0.5}$ &$9\pm7$\\
$Br(B^0\to D^{*+}_{s0}(2317)D^-)$ &$7.0^{+3.0+0.1+0.2}_{-2.1-0.2-0.3}$&$8.3^{+3.7+0.4+0.3}_{-2.4-0.2-0.3}$&$10.5^{+4.5+0.4+0.4}_{-3.0-0.2+0.4}$   &$10.6\pm1.6$\\
$Br(B^0\to D^{*+}_{s0}(2317)D^{*-})$    &$10.5^{+4.7+1.6+0.3}_{-3.2-0.9-0.4}$ &$12.6^{+5.7+1.9+0.5}_{-3.8-1.1-0.5}$&$15.9^{+7.0+2.4+0.6}_{-4.9-1.4-0.5}$&$15\pm6$\\
\hline\hline
\end{tabular}\label{tab2}
\end{center}
\end{table}

Certainly, here most uncertainty parameter is the decay constant
$f_{D^*_{s0}}$, which is defined by the matrix element of the vector
current \be \langle0|\bar s\gamma_{\mu}c|D^*_{s0}(\textbf{p})\rangle
=f_{D^*_{s0}}p_{\mu}. \en
It connects with another decay constant $\tilde{f}_{D^*_{s0}}$ defined in Eq.(\ref{barfDst0}) at zero momentum
by the formula
$f_{D^*_{s0}}=\tilde{f}_{D^*_{s0}}(m_c-m_s)/m_{D^*_{s0}}$. The decay
constant $f_{D^*_{s0}}$ has been computed by different approaches
with results covering a wide range (shown in Table \ref{tab1}). It
is interesting that the works \cite{cheng,cheng1,faessler,cheng3}
about the analysis of the decay constant $f_{D^*_{s0}}$ through B decays
to two charmed mesons are consistent with each
other:$f_{D^*_{s0}}$ is in the range of $60\sim75$MeV, which is much
smaller than the decay constant of another P-wave meson
$D_{s1}(2460)$. That is to say that there exists large disparity
between these two decay constants. And the corresponding analysis
approaches include the heavy quark symmetry (HQS), the light front
quark model (QM). While some authors considered that $f_{D^*_{s0}}$
is larger than 100 MeV by using the quark model \cite{hsieh}, lattice QCD (LQCD) \cite{bali} and so on.

From our calculations, we find that the smaller decay constant
$f_{D^*_{s0}}$ is supported by the present data, say $55\sim70$ MeV.
The values such as larger than $100$ MeV seem are not favored. So we
calculate the branching ratios corresponding to $f_{D^*_{s0}}=55,
60, 67$ MeV and list in Table \ref{tab2}.

It is helpful to define the following two ratios \be
R_1=\frac{Br(B^+\to D^{*+}_{s0}(2317)\bar D^0)}{Br(B^+\to
D^{*+}_{s0}(2317)\bar D^{*0})}),\;\;\; R_2=\frac{Br(B^0\to
D^{*+}_{s0}(2317)D^-)}{Br(B^0\to D^{*+}_{s0}(2317)D^{*-})}. \label{ratio2}\en They
are in the range $0.61\sim0.67$ (shown in Table \ref{tab3}), which
are lower than the experimental values $0.89$ and $0.71$,
respectively. Certainly, these two ratios can be well determined by
the future experiments through improving the measurement accuracy to
the decays $B^+\to D^{*+}_{s0}(2317)\bar D^{*0}$ and $B^0\to
D^{*+}_{s0}(2317)D^{*-}$. It is very possible that the ratios $R_1$ and
$R_2$ are less than 1, which is contrary to the previous prediction
\cite{cheng3}.

\begin{table}
\caption{The ratios $R_{1,2}$ defined in Eq.(\ref{ratio2}) with different values $f_{D^*_{s0}}=55, 60, 67$ MeV, where the
errors are the same with those in Tab.2, but with them added in quadrature.}
\begin{center}
\begin{tabular}{c|c|c|c|c}
\hline\hline Modes & $f_{D^*_{s0}}=55$ &  $f_{D^*_{s0}}=60$ & $f_{D^*_{s0}}=67$ & Data \cite{pdg20}  \\
\hline
$R_1$&$0.63^{+0.39}_{-0.27}$&$0.62^{+0.40}_{-0.26}$&$0.61^{+0.33}_{-0.24}$&$0.89^{+0.71}_{-0.70}$\\
$R_2$&$0.67^{+0.43}_{-0.29}$&$0.66^{+0.43}_{-0.28}$&$0.66^{+0.42}_{-0.28}$&$0.71\pm0.30$\\
\hline\hline
\end{tabular}\label{tab3}
\end{center}
\end{table}

Here we take $B^0\to D^{*+}_{s0}(2317)D^-$ as an example to do the numerical analysis about
the amplitudes from different types of Feynman diagrams. The tree
operators from the factorizable emission diagrams give the largest
contribution because of the large Wilson coefficient $C_2+C_1/3$,
and the value of the corresponding amplitude is about $4.39\times10^{-2}$. The
amplitude from the nonfactorizable emission diagrams is suppressed
by the small Wilson coefficient $C_1$, whose value is about
$(1.32+i0.57)\times10^{-2}$. The total amplitude of penguin operators is about
$-(0.95+i0.07)\times10^{-2}$, which comes from two parts: One is the factorizable
and nonfactorizable emission diagrams $-(0.93+i0.04)\times10^{-2}$, the other is the
annihilation diagrams $-(1.11+i3.67)\times10^{-4}$. The penguin
operators receive severe suppression from the Wilson coefficients
and only contribute $3.9\%$ to the final branching ratio. The
penguin operator contributions from the annihilation diagrams are even
tiny and can be neglected. So it is enough to pay our attention only
to the factorizable and nonfactorizable emission diagrams for the
investigation of the branching ratios. As to the branching ratios
for the decays of $B_s$ to two charmed mesons are also calculated
and listed in Table \ref{tab4}. In these decays, $B_s\to
D^{*-}_{s0}(2317)D^{(*)+}_s$ have the largest branching ratios,
which are at $10^{-3}$ order. They are consistent with the
predictions by the relativistic quark model (RQM) \cite{faus}, while
are much smaller than those by using the light cone sum rules (LCSR)
approach \cite{lirh}. It can be tested by the future LHCb and Super
KEKB experiments. For the decays $B_s\to
D^{*-}_{s0}(2317)D^{(*)+}$,their branching ratios are much smaller
than other four channels mainly because of the smaller CKM matrix
element $V_{cd}$ compared with $V_{cs}$, that is to say there
exists a suppressed factor $|V_{cd}/V_{cs}|^2\approx0.05$ between
the branching ratios of these two types of decays.


\begin{table}
\caption{Branching ratios ($\times10^{-3}$) of the decays $B_s\to
D^{*}_{s0}(2317)D^{(*)}_s,D^{*}_{s0}(2317)D^{(*)}$ with
different values $f_{D^*_{s0}}=55, 60, 67$ MeV, where the
errors for these entries correspond to the uncertainties in the
$w_{b}=0.5\pm0.05$ GeV for $B_s$ meson, the hard scale $t$ varying from
$0.75t$ to $1.25t$, and the CKM matrix elements. Some of these
channels have been calculated by using the light cone sum rules
(LCSR) \cite{lirh} and the relativistic quark model (RQM)
 \cite{faus}, which are listed in the last two columns. }
{\scriptsize
\begin{center}
\begin{tabular}{c|c|c|c|c|c}
\hline\hline Modes & $f_{D^*_{s0}}=55$MeV& $f_{D^*_{s0}}=60$MeV& $f_{D^*_{s0}}=67$MeV&  RQM \cite{faus} & LCSR \cite{lirh} \\
\hline
$B_s\to D^{*-}_{s0}(2317)D^+_s$       &$1.4^{+0.6+0.2+0.1}_{-0.4-0.1-0.0}$  &$1.7^{+0.8+0.2+0.0}_{-0.5-0.1-0.1}$  &$2.1^{+0.9+0.3+0.1}_{-0.6-0.1-0.1}$ &$1.1$&$13^{+7}_{-5}$\\
$B_s\to D^{*-}_{s0}(2317)D^{*+}_s$    &$1.2^{+0.5+0.1+0.0}_{-0.4-0.1-0.1}$  & $1.4^{+0.8+0.1+0.1}_{-0.4-0.1-0.0}$ &$1.8^{+0.9+0.1+0.1}_{-0.6-0.1-0.1}$              &$2.3$&$6.0^{+2.9}_{-2.4}$\\
$B_s\to D^{*+}_{s0}(2317)D^-_s$ &$0.73^{+0.35+0.02+0.03}_{-0.24-0.04-0.01}$ & $0.86^{+0.56+0.04+0.03}_{-0.27-0.03-0.03}$    & $1.11^{+0.56+0.02+0.04}_{-0.37-0.04-0.04}$ & & \\
$B_s\to D^{*+}_{s0}(2317)D^{*-}_s$    &$0.97^{+0.45+0.07+0.03}_{-0.31-0.06-0.04}$ &$1.17^{+0.56+0.08+0.04}_{-0.37-0.07-0.05}$  &$1.48^{+0.69+0.05+0.06}_{-0.46-0.07-0.05}$ & & \\
$B_s\to D^{*-}_{s0}(2317)D^+$    &$0.043^{+0.023+0.004+0.002}_{-0.014-0.003-0.001}$     & $0.052^{+0.027+0.005+0.002}_{-0.017-0.003-0.002}$ &$0.065^{+0.034+0.006+0.002}_{-0.021-0.004-0.002}$ & &$0.5^{+0.2}_{-0.2}$\\
$B_s\to D^{*-}_{s0}(2317)D^{*+}$ &$0.033^{+0.018+0.003+0.001}_{-0.011-0.002-0.001}$    &$0.040^{+0.021+0.003+0.001}_{-0.014-0.002-0.001}$  &$0.050^{+0.026+0.004+0.002}_{-0.017-0.002-0.001}$ & &$0.2^{+0.1}_{-0.1}$\\
\hline\hline
\end{tabular}\label{tab4}
\end{center}}
\end{table}
Through our calculations, we find that the direct CP asymmtries of
our considered decays are in the range of $10^{-4}\sim10^{-3}$. For
example, $A^{dir}_{CP}(B^+\to D^+_{s0}(2317)D^{0})$ is about
$0.75\%$. As we know that the direct CP asymmetry is proportional to
the interference between the tree and penguin contributions, while
the penguin contributions are small as we mentioned above, so it is
not surprise that the direct CP violation of these decays are small.
In a word, the direct CP asymmetries in these decays of $B$ to
two charmed mesons are tiny, any large direct CP violation
observed in the future experiments can be treated as a new dynamic
mechanism from the some special structure of $D^{*+}_{s0}(2317)$ or
a signal of new physics.
\section{conclusion}
In summary, we probe the inner structure of the meson
$D^*_{s0}(2317)$ through the decays of $B_{(s)}$ to two charmed
mesons within pQCD approach. Assuming $D^*_{s0}(2317)$ as a scalar
meson with $\bar cs$ structure, we find that our predictions for the
branching ratios of the decays $B^+\to D^{*+}_{s0}(2317)\bar
D^{(*)0}, B^0\to D^{*+}_{s0}(2317)D^{(*)-}$ can explain data within
errors. In our calculations, the decay constant of the meson
$D^*_{s0}(2317)$ is an input parameter, and its value has
been studied by many references but with results covering a wide
range. While a smaller value of the decay constant for the meson
$D^*_{s0}(2317)$ is supported in our work, say $55\sim70$
MeV. We also calculate the ratios $R_1=\frac{Br(B^+\to
D^{*+}_{s0}(2317)\bar D^0)}{Br(B^+\to D^{*+}_{s0}(2317)\bar
D^{*0})})$ and $R_2=\frac{Br(B^0\to D^{*+}_{s0}(2317)D^-)}{Br(B^0\to
D^{*+}_{s0}(2317)D^{*-})}$, which are valuable to determine the
inner structure of the meson $D^*_{s0}(2317)$ by compared between theory and
experiment. We expect that these two values can be well measured by
the future LHCb and Super KEKB experiments through improving the
measurement accuracy for the decays $B^+\to D^{*+}_{s0}(2317)\bar
D^{*0}$ and $B^0\to D^{*+}_{s0}(2317)D^{*-}$.
\section*{Acknowledgment}
This work is supported by the National Natural Science Foundation of
China under Grant No. 11347030, 11847097 and by the Program of Science and
Technology Innovation Talents in Universities of Henan Province
14HASTIT037. One of us (N. Wang) is supported by the Science Research Fund Project for the High-Level Talents of
Henan University of Technology 004/31401151.
\appendix
\appendix
\section{Scales, functions for the hard kernel and The evolution factors}
The variables that are evaluated from the gluon and quark
propagators will be used to determine the scales and the expressions of
the hard kernels \be
P_{en}&=&m^2_Bx_1x_3(1-r^2_{D^*_{s0}}), P^{(1)}_{en}=m^2_Bx_3(x_1(1-r^2_{D^*_{s0}})-x_2(1-r^2_{D^*_{s0}}-r^2_{D_s})),\label{pen}\\
P^{(2)}_{en}&=&m^2_B\left[r^2_c-((1-x_1-x_2)x_3-(1-x_2)x_3r^2_{D_s}+(1-x_1-x_2)(1-x_3)r^2_{D^*_{s0}})\right],\\
P_{an}&=&-m^2_B\left[1-(1-r^2_{D_s})x_2-(1-r^2_{D^*_{s0}})x_3+x_2x_3(1-r^2_{D_s}-r^2_{D^*_{s0}})\right],\\
P^{(1)}_{an}&=&m^2_B\left[1+(1-r^2_{D^*_{s0}})x_1x_3-(1-r^2_{D_s}-r^2_{D^*_{s0}})x_2x_3\right],\\
P^{(2)}_{an}&=&m^2_B\left[x_1+x_2+x_3-1-x_1x_3(1-r^2_{D^*_{s0}})-x_2r^2_D-x_3r^2_{D^*_{s0}}-x_2x_3(1-r^2_D-r^2_{D^*_{s0}})\right],\non\\
P_{ancs}&=&-m^2_B\left(1-r^2_{D_s}-r^2_{D^*_{s0}}\right)x_2x_3,P^{(2)}_{ancs}=m^2_B\left[x_1(1-r^2_{D^*_{s0}})-(1-r^2_{D_s}-r^2_{D^*_{s0}})x_2\right]x_3,\non\\
P^{(1)}_{ancs}&=&m^2_B\left[x_1(1+(r^2_{D^*_{s0}}-1)x_3)+x_3(1-r^2_{D^*_{s0}})+x_2(1-x_3(1-r^2_{D^*_{s0}})+(x_3-1)r^2_{D_s})\right].\non
\en Then the scales in each amplitude are determined as \be
t^{(1)}_{e}&=&\max(\sqrt{x_3(1-r^2_{D^*_{s0}})}m_B,1/b_1,1/b_3),\\
t^{(2)}_e&=&\max(\sqrt{x_1(1-r^2_{D^*_{s0}})}m_B,1/b_1,1/b_3),\\
t^{(1,2)}_{en}&=&\max(\sqrt{P_{en}},\sqrt{|P^{(1,2)}_{en}|},1/b_1,1/b_2),\\
t^{(1,2)}_{an}&=&\max(\sqrt{|P_{an}|},\sqrt{|P^{(1,2)}_{an}|},1/b_1,1/b_3),\\
t^{(1,2)}_{ancs}&=&\max(\sqrt{|P_{ancs}|},\sqrt{|P^{(1,2)}_{ancs}|},1/b_1,1/b_2),\\
t^{(1)}_a&=&\max(\sqrt{1-(1-r^2_{D^*_{s0}})x_3},1/b_2,b_3),\\
t^{(2)}_a&=&\max(\sqrt{1-(1-r^2_{D_{s}})x_2},1/b_2,b_3),\\
t^{(1)}_{acs}&=&\max(\sqrt{(1-r^2_{D_{s}}-r^2_{D^*_{s0}})x_2},1/b_2,b_3),\\
t^{(2)}_{acs}&=&\max(\sqrt{(1-r^2_{D_{s}}-r^2_{D^*_{s0}})x_3},1/b_2,b_3).
\en The hard functions for the hard part of the amplitudes are
listed as \be
h_e(x_1,x_3,b_1,b_3)&=&K_0(\sqrt{x_1x_3}m_{B_s}b_1)\left[\theta(b_1-b_3)K_0(\sqrt{x_3}m_{B_s}b_1)
I_0(\sqrt{x_3}m_{B_s}b_3)\right.
\non && \left.+\theta(b_3-b_1)K_0(\sqrt{x_3}m_{B_s}b_3)I_0(\sqrt{x_3}m_{B_s}b_1)\right],\\
h_a(x_2,x_3,b_2,b_3)&=&\left(i\frac{\pi}{2}\right)^2H^{(1)}_0(\sqrt{x_2x_3}m_{B_s}b_2)\left[\theta(b_2-b_3)H^{(1)}_0(\sqrt{x_3}m_{B_s}b_2)J_0(\sqrt{x_3}m_{B_s}b_3)\right.\non
&&+\left.\theta(b_3-b_2)H^{(1)}_0(\sqrt{x_3}m_{B_s}b_3)J_0(\sqrt{x_3}m_{B_s}b_2)\right],\\
h^{(j)}_{en}(x_1,x_2,x_3,b_1,b_2)&=&\left[\theta(b_1-b_2)K_0(\sqrt{P_{en}}b_1)
I_0(\sqrt{P_{en}}b_2)\right.\non && \left.+(b_1\leftrightarrow
b_2)\right]
\left(\begin{matrix}K_0(\sqrt{P^{(j)}_{en}}b_2)& \text{for} P^{(j)}_{en}\geq 0\\
\frac{i\pi}{2}H^{(1)}_0(\sqrt{|P^{(j)}_{en}|}b_2)&\text{for}
P^{(j)}_{en}\leq
0\\\end{matrix}\right),\\
h^{(j)}_{an(cs)}(x_1,x_2,x_3,b_1,b_3)&=&i\frac{\pi}{2}\left[\theta(b_1-b_3)H^{(1)}_0(\sqrt{P_{an(cs)}}b_1)
J_0(\sqrt{P_{an(cs)}}b_3)\right.\non && \left.+(b_1\leftrightarrow
b_3)\right]
\left(\begin{matrix}K_0(\sqrt{P^{(j)}_{an(cs)}}b_1)& \text{for} P^{(j)}_{an(cs)}\geq 0\\
\frac{i\pi}{2}H^{(1)}_0(\sqrt{|P^{(j)}_{an(cs)}|}b_1)&\text{for}
P^{(j)}_{an(cs)}\leq 0\\\end{matrix}\right), \en where the functions
$H^{(1)}_0,J_0, K_0, I_0$ are the (modified) Bessel functions and
obtained from the Fourier transformations of the quark and gluon
propagators. The evolution factors evolving the scale $t$ are
defined as \be
E_e(t)&=&\alpha_s(t)\exp[-S_{B_s}(t)-S_{D_{s}}(t)],\label{suda1}\\
E_{en}(t)&=&\alpha_s(t)\exp[-S_{B_s}(t)-S_{D_s}(t)-S_{D^*_{s0}}(t)|_{b_1=b_3}],\label{suda2}\\
E_{an}(t)&=&\alpha_s(t)\exp[-S_{B_s}(t)-S_{D_s}(t)-S_{D^*_{s0}}(t)|_{b_2=b_3}],\label{suda3}\\
E_{a}(t)&=&\alpha_s(t)\exp[-S_{D_s}(t)-S_{D^*_{s0}}(t)],\label{suda4}
\en where the definitions of the functions $S_j(t)
(j={B_s},D_{D^*_{s0}},D_s)$ in Eq.(\ref{suda1}), Eq.(\ref{suda2}),
Eq.(\ref{suda3}) and Eq.(\ref{suda4}) are given as \be
S_{B_s}(t)&=&s(x_1\frac{m_{B_s}}{\sqrt2},b_1)+\frac{5}{3}\int^t_{1/b_1}\frac{d\bar\mu}{\bar\mu}\gamma_q(\alpha_s(\bar\mu)),\\
S_{D_s}(t)&=&s(x_2\frac{m_{B_s}}{\sqrt2},b_2)+2\int^t_{1/b_2}\frac{d\bar\mu}{\bar\mu}\gamma_q(\alpha_s(\bar\mu)),\\
S_{D^*_{s0}}(t)&=&s(x_3\frac{m_{B_s}}{\sqrt2},b_3)+2\int^t_{1/b_3}\frac{d\bar\mu}{\bar\mu}\gamma_q(\alpha_s(\bar\mu)).
\en Here the quark anomalous dimension $\gamma_q=-\alpha_s/\pi$,
and the expression of the $s(Q,b)$ in one-loop running coupling
coupling constant is used \be
s(Q,b)&=&\frac{A^{(1)}}{2\beta_1}\hat{q}\ln(\frac{\hat{q}}{\hat{b}})-\frac{A^{(1)}}{2\beta_1}(\hat{q}-\hat{b})
+\frac{A^{(2)}}{4\beta^2_1}(\frac{\hat{q}}{\hat{b}}-1)\non
&&-\left[\frac{A^{(2)}}{4\beta^2_1}-\frac{A^{(1)}}{4\beta_1}\ln(\frac{e^{2\gamma_E-1}}{2})\right]
\ln(\frac{\hat{q}}{\hat{b}}), \en with the variables are defined by
$\hat{q}=\ln[Q/(\sqrt2\Lambda)], \hat{q}=\ln[1/(b\Lambda)]$ and the
coefficients $A^{(1,2)}$ and $\beta_{1}$ are \be
\beta_1&=&\frac{33-2n_f}{12},A^{(1)}=\frac{4}{3},\\
A^{(2)}&=&\frac{67}{9}-\frac{\pi^2}{3}
-\frac{10}{27}n_f+\frac{8}{3}\beta_1\ln(\frac{1}{2}e^{\gamma_E}),
\en where $n_f$ is the number of the quark flavors and $\gamma_E$ the
Euler constant.
\section{Amplitudes for the decay $B^0_s\to D^{*-}_{s0}(2317)D^+_s$}
The amplitudes for the decay $B^0_s\to D^{*-}_{s0}(2317)D^+_s$ are
listed in the following. It is noticed that we add the character "c"
in the subscripts to distinguish them from those for Fig.1, which
represents a conventional charmed meson $D_s$ being the emission
(upper) position in the Feynman diagrams. \be
\mathcal{F}^{LL}_{ec}&=&-\mathcal{F}^{LR}_{ec}=8\pi
C_FM^4_{B_s}f_{D_s}\int_0^1 d x_{1} dx_{2}\, \int_{0}^{\infty} b_1
db_1 b_2 db_2\, \phi_{B_s}(x_1,b_1)\phi_{D^*_{s0}}(x_2)\non &&
\times [1+r_{D^*_{s0}}+(1-2r_{D^*_{s0}})x_2]
E_{ec}(t^{(1)}_{ec})S_t(x_2)h_e(x_1,x_2(1-r^2_{D_{s}}),b_1,b_2)\non
&&
+\left[r_c(1-2r_{D^*_{s0}})+r_{D^*_{s0}}(2-r_{D^*_{s0}})\right]E_{ec}(t^{(2)}_{ec})S_t(x_1)
h_e(x_2,x_1(1-r^2_{D_{s}}),b_2,b_1)], \label{fec1}\\
 \mathcal{F}^{SP}_{ec}&=&-16\pi
C_FM^4_{{B_s}}f_{D_s}r_{{D_s}}\int_0^1 d x_{1} dx_{2}\,
\int_{0}^{\infty} b_1 db_1 b_2 db_2\, \phi_{B_s}(x_1,b_1)\non &&
\times
\left\{[1+r_{D^*_{s0}}(2+r_{D^*_{s0}}+x_2(1-4r_{D^*_{s0}}))]E_{ec}(t^{(1)}_{ec})S_t(x_2)h_e(x_1,x_2(1-r^2_{D_s}),b_1,b_2)\right.\non
&&\left.
+[r_c(1-4r_{D^*_{s0}})+2r_{D^*_{s0}}(1-r_{D^*_{s0}})]E_{ec}(t^{(2)}_{ec})S_t(x_1)
h_e(x_2,x_1(1-r^2_{D_s}),b_2,b_1)\right\}, \;\;\;\;\;\;\label{fec2}\\
\mathcal{M}^{LL}_{enc}&=&32\pi C_f m_{B_s}^4/\sqrt{2N_C}\int_0^1 d
x_{1} dx_{2} dx_{3}\, \int_{0}^{\infty} b_1 db_1 b_3
db_3\,\phi_{B_s}(x_1,b_1)\phi_{D^*_{s0}}(x_2)\phi_{D_s}(x_3)\non &&
\times\left\{\left[x_3-r_{D^*_{s0}}x_2(1-2r_{D^*_{s0}})\right]E_{enc}(t^{(1)}_{enc})h^{(1)}_{enc}(x_1,x_3,x_2,b_1,b_3)
\right.\non &&\left.
 +[x_3-1-(1-r_{D^*_{s0}})x_2+r_cr_D]E_{enc}(t^{(2)}_{enc})h^{(2)}_{enc}(x_1,x_3,x_2,b_1,b_3)\right\},
\en \be \mathcal{M}^{LR}_{enc}&=&-32\pi C_f
m_{B_s}^4/\sqrt{2N_C}\int_0^1 d x_{1} dx_{2} dx_{3}\,
\int_{0}^{\infty} b_1 db_1 b_3
db_3\,\phi_{B_s}(x_1,b_1)\phi_{D^*_{s0}}(x_2)\phi_{{D_s}}(x_3)\non
&&
\times(r_{D^*_{s0}}+1)\left\{r_{D_s}\left[x_3+r_{D^*_{s0}}(1+r_{D^*_{s0}})x_2\right]E_{enc}(t^{(1)}_{enc})h^{(1)}_{enc}(x_1,x_3,x_2,b_1,b_3)
\right.\non &&\left.
 -[r_c+r_{D_s}(1+r_{D^*_{s0}})(1-x_3)]E_{enc}(t^{(2)}_{enc})h^{(2)}_{enc}(x_1,x_3,x_2,b_1,b_3)\right\},\\
\mathcal{M}^{SP}_{enc}&=&-32\pi C_f m_{B_s}^4/\sqrt{2N_C}\int_0^1 d
x_{1} dx_{2} dx_{3}\, \int_{0}^{\infty} b_1 db_1 b_3
db_3\,\phi_B(x_1,b_1)\phi_{D^*_{s0}}(x_2)\phi_{D_s}(x_3)\non &&
\times\left\{\left[x_3+(1-r_{D^*_{s0}})x_2\right]E_{enc}(t^{(1)}_{enc})h^{(1)}_{enc}(x_1,x_3,x_2,b_1,b_3)
\right.\non &&\left.
 +[x_3+r_{D^*_{s0}}x_2-1]E_{enc}(t^{(2)}_{enc})h^{(2)}_{enc}(x_1,x_3,x_2,b_1,b_3)\right\},
\en where these amplitudes are factorizable and nonfactorizable
emission contributions, respectively. The nonfactorizable and
factorizable annihilation amplitudes are written as \be
\mathcal{M}^{LL}_{anc}&=&32\pi C_f m_{B_s}^4/\sqrt{2N_C}\int_0^1 d
x_{1} dx_{2} dx_{3}\,\int_{0}^{\infty} b_1 db_1 b_2 db_2\,
\phi_{B_s}(x_1,b_1)\phi_{D^*_{s0}}(x_2)\phi_{{D_s}}(x_3)\non &&
\times\left\{\left[x_3-1-r_{D_s}r_{D^*_{s0}}(x_2+x_3-4)\right]
E_{an}(t^{(1)}_{anc})h^{(1)}_{anc}(x_1,x_3,x_2,b_1,b_2)\right.\non &&\left.+\left[1-x_2+r_{D_s}r_{D^*_{S0}}(x_2+x_3-2)\right]E_{an}(t^{(2)}_{anc})h^{(2)}_{anc}(x_1,x_3,x_2,b_1,b_2)\right\},\\
\mathcal{M}^{LR}_{anc}&=&-32\pi C_f m_B^4/\sqrt{2N_C}\int_0^1 d
x_{1} dx_{2} dx_{3}\,\int_{0}^{\infty} b_1 db_1 b_2 db_2\,
\phi_{B_s}(x_1,b_1)\phi_{D^*_{s0}}(x_2)\phi_{D_s}(x_3)\non &&
\times\left\{\left[r_{D_s}(1+x_3)+r_{D^*_{s0}}(1+x_2)\right]
E_{an}(t^{(1)}_{anc})h^{(1)}_{anc}(x_1,x_3,x_2,b_1,b_2)\right.\non &&\left.+\left[r_{D^*_{s0}}(1-x_2)+r_{D_s}(1-x_3)\right]E_{an}(t^{(2)}_{anc})h^{(2)}_{anc}(x_1,x_3,x_2,b_1,b_2)\right\},\\
\mathcal{M}^{SP}_{anc}&=&32\pi C_f m_B^4/\sqrt{2N_C}\int_0^1 d x_{1}
dx_{2} dx_{3}\,\int_{0}^{\infty} b_1 db_1 b_2 db_2\,
\phi_{B_s}(x_1,b_1)\phi_{D^*_{s0}}(x_2)\phi_{D_s}(x_3)\non &&
\times\left\{\left[1-x_2+r_{D_s}r_{D^*_{s0}}(x_2+x_3-4)\right]
E_{an}(t^{(1)}_{anc})h^{(1)}_{anc}(x_1,x_3,x_2,b_1,b_2)\right.\non
&&\left.-\left[1-x_3+r_{D_s}r_{D^*_{s0}}(x_2+x_3-2)\right]E_{an}(t^{(2)}_{anc})h^{(2)}_{anc}(x_1,x_3,x_2,b_1,b_2)\right\},
\en \be \mathcal{F}^{LL}_{ac}&=&-\mathcal{F}^{LR}_{ac}=8\pi C_f
f_{B_s}m_{B_s}^4\int_0^1 d x_{2} dx_{3}\, \int_{0}^{\infty} b_2 db_2
b_3 db_3\, \phi_{D^*_{s0}}(x_2)\phi_{{D_s}}(x_3)\non
&&\left\{\left[x_2-1-2r_{D_s}r_{D^*_{s0}}(x_2-2)\right]
E_{a}(t^{(1)}_{ac})S_t(x_2)h_{a}(\overline{(1-r_{D^*_{s0}}^2)x_3},\overline{(1-r^2_{D_s})x_2},b_3,b_2)
\right.\non
&&\left.+\left[1-x_3+2r_{D_s}r_{D^*_{s0}}(x_3-2)\right]E_{a}(t^{(2)}_{ac})S_t(x_3)h_{a}(\overline{(1-r^2_{D_s})x_2},\overline{(1-r_{D^*_{S0}}^2)x_3},b_2,b_3)\right\},\non
\\
\mathcal{F}^{SP}_{ac}&=&16\pi C_f f_{B_s}m_{B_s}^4\int_0^1 d x_{2}
dx_{3}\, \int_{0}^{\infty} b_2 db_2 b_3 db_3\,
\phi_{D^*_{s0}}(x_2)\phi_{D_s}(x_3)\non
&&\left\{\left[-2r_{D_s}+r_{D^*_{s0}}(1-x_2)\right]
E_{a}(t^{(1)}_{ac})S_t(x_2)h_{a}(\overline{(1-r_{D^*_{s0}}^2)x_3},\overline{(1-r^2_{D_s})x_2},b_3,b_2)
\right.\non &&\left.+
\left[2r_{D^*_{s0}}-r_{D_s}(1-x_3)\right]E_{a}(t^{(2)}_{ac})S_t(x_3)h_{a}(\overline{(1-r^2_{D_s})x_2},\overline{(1-r_{D^*_{s0}}^2)x_3},b_2,b_3)\right\}.
\en If changing the the positions of the scalar and the pseudoscalar
mesons in the final states, one can get another type of annihilation
amplitudes, which are given as \be
\mathcal{M}^{LL}_{ans}&=&-32\pi C_f m_{B_s}^4/\sqrt{2N_C}\int_0^1 d
x_{1} dx_{2} dx_{3}\,\int_{0}^{\infty} b_1 db_1 b_3 db_3\,
\phi_{B_s}(x_1,b_1)\phi_{D^*_{s0}}(x_2)\phi_{D_s}(x_3)\non &&
\times\left\{\left[x_2+r_{D_s}r_{D^*_{s0}}(x_2+x_3+2)\right]
E_{an}(t^{(1)}_{ans})h^{(1)}_{ans}(x_1,x_2,x_3,b_1,b_3)\right.\non &&\left.-\left[x_3+r_{D_s}r_{D^*_{s0}}(x_2+x_3)\right]E_{an}(t^{(2)}_{ans})h^{(2)}_{ans}(x_1,x_3,x_2,b_1,b_3)\right\},\\
\mathcal{M}^{LR}_{ans}&=&-32\pi C_f m_{B_s}^4/\sqrt{2N_C}\int_0^1 d
x_{1} dx_{2} dx_{3}\,\int_{0}^{\infty} b_1 db_1 b_3 db_3\,
\phi_{B_s}(x_1,b_1)\phi_{D^*_{s0}}(x_2)\phi_{D_s}(x_3)\non &&
\times\left\{\left[r_{D_s}(2-x_3)+r_{D^*_{s0}}(x_2-2)\right]
E_{an}(t^{(1)}_{ans})h^{(1)}_{ans}(x_1,x_2,x_3,b_1,b_3)\right.\non &&\left.-\left[r_{D^*_{s0}}x_2-r_{D_s}x_3\right]E_{an}(t^{(2)}_{ans})h^{(2)}_{ans}(x_1,x_2,x_3,b_1,b_3)\right\},
\en\be
\mathcal{M}^{SP}_{ans}&=&32\pi C_f m_{B_s}^4/\sqrt{2N_C}\int_0^1 d
x_{1} dx_{2} dx_{3}\,\int_{0}^{\infty} b_1 db_1 b_3 db_3\,
\phi_{B_s}(x_1,b_1)\phi_{D^*_{s0}}(x_2)\phi_{D_s}(x_3)\non &&
\times\left\{\left[x_3+r_{D_s}r_{D^*_{s0}}(x_2+x_3+2)\right]
E_{an}(t^{(1)}_{ans})h^{(1)}_{ans}(x_1,x_2,x_3,b_1,b_3)\right.\non
&&\left.-\left[x_2+r_{D_s}r_{D^*_{s0}}(x_2+x_3)\right]E_{an}(t^{(2)}_{ans})h^{(2)}_{ans}(x_1,x_2,x_3,b_1,b_3)\right\},
\en \be \mathcal{F}^{LL}_{as}&=&-\mathcal{F}^{LR}_{as}=-8\pi C_f
f_{B_s}m_{B_s}^4\int_0^1 d x_{2} dx_{3}\, \int_{0}^{\infty} b_2 db_2
b_3 db_3\, \phi_{D^*_{s0}}(x_2)\phi_{D_s}(x_3)\non
&&\left\{\left[x_3+2r_{D_s}r_{D^*_{s0}}(1+x_3))\right]
E_{a}(t^{(1)}_{as})S_t(x_3)h_{a}(x_2,(1-r_{D^*_{s0}}^2-r^2_{D_s})x_3,b_2,b_3))
\right.\non &&\left.-\left[x_2+2r_{D_s}r_{D^*_{s0}}(x_2+1)\right]E_{a}(t^{(2)}_{as})S_t(x_2)h_{a}(x_3,(1-r_{D^*_{S0}}^2-r^2_{D_s})x_2,b_3,b_2))\right\},\non\label{ckms4}\\
\mathcal{F}^{SP}_{as}&=&16\pi C_f f_{B_s}m_{B_s}^4\int_0^1 d x_{2}
dx_{3}\, \int_{0}^{\infty} b_2 db_2 b_3 db_3\,
\phi_{D^*_{s0}}(x_2)\phi_{D_s}(x_3)\non
&&\left\{\left[2r_{D^*_{s0}}+r_{{D_s}}x_3+r_c\right]
E_{a}(t^{(1)}_{as})S_t(x_3)h_{a}(x_2,(1-r_{D^*_{s0}}^2-r^2_{D_s})x_3,b_2,b_3))
\right.\non &&\left.+
\left[2r_{{D_s}}+r_{D^*_{s0}}x_2-r_c\right]E_{a}(t^{(2)}_{as})S_t(x_2)h_{a}(x_3,(1-r_{D^*_{s0}}^2-r^2_{D_s})x_2,b_2,b_3))\right\}.
\en From the previews contents, it is easy to know that the
character "s" in each subscript represents $s\bar s$ pair generated
from a hard gluon in the corresponding Feynman diagrams. The scales,
the functions for the hard kernel and the evolution factors can be
obtained for Eq.(\ref{pen}) to Eq.(\ref{suda4}) by the following
substitutions \be x_2\leftrightarrow x_3, b_2\leftrightarrow b_3,
r_{D^*_{s0}}\leftrightarrow r_{D_s}. \en

\end{document}